\documentclass[a4paper,fleqn,usenatbib,useAMS]{mnras}

\usepackage{graphicx}	
\usepackage{amsmath}	
\usepackage{amssymb}	
\usepackage{multicol}       
\usepackage{bm}		
\usepackage{pdflscape}	
\usepackage{color}
\usepackage{soul}

\newcommand{\kms}{\,km\,s$^{-1}$} 
\newcommand\error{$\pm$}

\usepackage[T1]{fontenc}
\usepackage{ae,aecompl}

\title[New members and halo stars of the globular cluster NGC~1851]
{New cluster members and halo stars of the Galactic globular cluster NGC~1851}

\author[C.A. Navin et al.]
  {Colin~A.~Navin$^1$,
   Sarah~L.~Martell$^2,$ and Daniel~B.~Zucker$^{1,3}$
\\
$^1$Department of Physics \& Astronomy, Macquarie University, Sydney 2109, Australia\\
$^2$Department of Optics \& Astrophysics, University of New South Wales, Sydney 2052, Australia\\
$^3$Australian Astronomical Observatory, PO Box 296, Epping, NSW 2121, Australia\\} 

\date{Last updated 2015 Jun 1; in original form 2015 Jun 1}

\pubyear{2015}

\begin{document}
\label{firstpage}
\pagerange{\pageref{firstpage}--\pageref{lastpage}}
\maketitle

\begin{abstract}
NGC~1851 is an intriguing Galactic globular cluster, with multiple stellar evolutionary sequences, light and heavy element abundance variations and indications of a surrounding stellar halo. We present the first results of a spectroscopic study of red giant stars within and outside of the tidal radius of this cluster. Our results identify nine probable new cluster members (inside the tidal radius) with heliocentric radial velocities consistent with that of NGC~1851. We also identify, based on their radial velocities, four probable extratidal cluster halo stars at distances up to $\sim$3.1 times the tidal radius, which are supportive of previous findings that NGC~1851 is surrounded by an extended stellar halo. Proper motions were available for 12 of these 13 stars and all are consistent with that of NGC~1851. Apart from the cluster members and cluster halo stars, our observed radial velocity distribution agrees with the expected distribution from a Besan\c con disk/N-body stellar halo Milky Way model generated by the Galaxia code, suggesting that no other structures at different radial velocities are present in our field. The metallicities of these stars are estimated using equivalent width measurements of the near infrared calcium triplet absorption lines and are found, within the limitations of this method, to be consistent with that of NGC~1851. In addition we recover 110 red giant cluster members from previous studies based on their radial velocities and identify three stars with unusually high radial velocities.
\end{abstract}




\section{Introduction}
\label{intro}

Until relatively recently globular clusters (GCs) were considered as examplars of single stellar populations (SSPs), with all the cluster stars having the same age, helium abundance and overall [Fe/H] metallicity. In reality all GCs that have been thoroughly studied have multiple stellar populations (with the possible exception of Ruprecht 106, but the sample size was small\citealt{Villanova:2013fj}). Evidence comes from many photometric studies showing multiple branches on colour-magnitude diagrams (CMDs), and high resolution spectroscopic studies showing that all globular clusters exhibit significant star-to-star abundance variations in light elements (such as C, N, O and Na). In addition a few clusters (typically the most massive) also show overall [Fe/H] metallicity variations and variations in neutron capture elements. 

Globular clusters are, therefore, more complex then originally thought, and in fact contain multiple generations of stars (for a review see \citealt{Gratton:2012lr}). This was first discovered for the more massive globular clusters and for the brightest red giant branch (RGB) and asymptotic giant branch (AGB) sequences. Now many globular clusters with a wide range of masses have been shown to exhibit these features, and various studies have shown multiple sequences in the fainter main sequences (MS), sub giant branches (SGB) and horizontal branches (HB). 

The abundances of light elements of stars in GCs exhibit various relations, in particular C-N and O-Na anticorrelations. The origin of these patterns is believed to be hot hydrogen burning in either AGB stars, fast rotating massive stars, massive binaries or supermassive stars \citep{Gratton:2012lr}. However there are indications that the C-N and O-Na anti-correlations are not perfectly coupled. \citet{Pancino:2010fk} measured CN and CH bandstrengths for MS stars in 12 GCs with well-characterized RGB abundances and found that the frequency of C-poor, N-rich MS stars does not match the frequency of O-poor, Na-rich RGB stars reported for those same clusters in \citet{Carretta:2009lr}. They suggested that enrichment in C and N may not match enrichment in O and Na because the CN cycle, the full CNO bi-cycle and the Ne-Na cycle are each responsible for a different part of the light element abundance pattern, and each requires a different temperature to operate.

NGC~1851 is an unusual Galactic globular cluster. It is one of four globular clusters (along with NGC 1904, NGC 2298 and NGC 2808) all physically located within a sphere of radius 6 kpc \citep{Bellazzini:2004fk}. It has been proposed that they form a group associated with the putative Canis Major dwarf galaxy stream \citep{Martin:2004lr}. NGC~1851 exhibits the common features of multiple evolutionary sequences and star-to-star abundance dispersions in the light elements, but it also has some features not ordinarily found in other globular clusters, as detailed below.

\subsection{Photometry}
\label{photometry}
\citet{Stetson:1981lr} detected peculiarities in the HB with a dense clump of stars at the red end and a hint of another clump at the blue end. \citet{Walker:1992fk} confirmed the split nature of the HB in the cluster, clearly shown in the right panel of Fig. \ref{cmd_combined}. A double SGB was announced in \citet{Milone:2008fk}; they postulated two populations with similar overall [Fe/H] metallicity but an age difference of $\sim$1 Gyr, while \citet{Cassisi:2008fk} suggested as an alternative two coeval populations with different C+N+O abundance mixtures. \citet{Han:2009qy} also detected a split RGB using UI photometry. These studies strongly suggest that a second generation of higher overall [Fe/H] metallicity stars formed after an early episode of metal enrichment.

\subsection{Spatial extent}
\label{spatial}

There have been a number of photometric and spectroscopic studies (detailed below) that indicate that NGC~1851 has a surrounding stellar halo outside the tidal radius. The nature of this halo is not yet clear - they may be stars stripped by tidal or disk shocking. Alternatively, modelling has shown that it is possible that NGC~1851 is the nuclear star cluster of a now disrupted dwarf galaxy, with the halo stars representing the remnant of that galaxy. 

The tidal radius $r_t$ calculated by different dynamical models such as King \citep{King:1966fk}, Wilson \citep{Wilson:1975lr} and power-law templates can vary considerably, and various studies have adopted different values. The current version of the {\citealt{Harris:1996lr}\ (2010 edition) catalogue does not tabulate $r_t$ values, but instead lists the structural parameters core radius $r_c$ = 0.09~arcmin and central concentration $c$ = 1.86~arcmin from \citet{McLaughlin:2005qy}. The central concentration $c$ is given by:
\begin{equation}
{c} = {log{\frac{r_t}{r_c}}}
\end{equation}
From this the tidal radius $r_t$ is calculated as 6.5~arcmin. The \citet{Wilson:1975lr} model $r_t$ as calculated by \citet{McLaughlin:2005qy}\ is 44.7~arcmin. \citet{Carballo-Bello:2012lm} looked at the number density profiles of 19 Galactic globular clusters and found power-law templates to be a better representation for about 2/3 of their sample, including NGC~1851 for which their calculated tidal radius is 11.7~arcmin. \citet{Allen:2006uq} calculated the Jacobi (or Roche) radius as 17~arcmin, although this is strongly dependent on the adopted Galaxy gravitational potential and cluster orbital parameters. For results discussed in this introduction, we quote the tidal radius adopted by the authors in the paper where applicable. For our own study we adopt the value of \citet{Carballo-Bello:2012lm} of 11.7~arcmin.

Wide field photometric star-count analyses carried out by \citet{Leon:2000qy} show a symmetric stellar structure out to $\sim$20~arcmin and extensions to $\sim$45~arcmin that are roughly aligned towards the Galactic center. Photometric studies by \citet{Olszewski:2009lr} also show indications of a stellar halo extending up to 75~arcmin, $\sim$6.5 times beyond their adopted tidal radius of 11.7~arcmin. This halo has a mass of $\sim$1~per~cent of the dynamical mass of NGC~1851, but there was no evidence for any substructure or tidal stream formation. \citet{Carballo-Bello:2010lr} reported a distinct MS of a metal-poor stellar overdensity with a sky projected size $>$15.7\degr~in the area surrounding NGC~1851 and NGC 1904.

A possible tidal tail was reported by \citet{Sollima:2012uq} in a spectroscopic survey of 107 stars from 12~arcmin to 33~arcmin around the cluster centre. They not only found a significant fraction of stars outside their adopted tidal radius of 11.7~arcmin, but also an interesting hint of a kinematically cold stellar stream, albeit with a peak in the heliocentric radial velocity ($V_r$) distribution of $\sim$180~\kms, compared with 320.5 \error\ 0.6~\kms~\citealt{Harris:1996lr} (2010 edition) for NGC~1851. \citet{Marino:2014lr} carried out a spectroscopic analysis of 23 stars in the halo of NGC~1851 and found stars at distances up to $\sim$2.5 times their adopted tidal radius of 11.7~arcmin. However they did not find convincing evidence for the tidal tail of stars reported by \citet{Sollima:2012uq} at $\sim$180~\kms. \citet{Kunder:2014lr} reported nine RAVE stars up to 10\degr~away from NGC~1851 that may be associated with the cluster.

 \citet{Bekki:2012qy} simulated the dynamical evolution of a dwarf galaxy with a stellar nucleus in the tidal field of the Galaxy. They showed that a diffuse stellar halo around NGC~1851 would be produced after stripping the dark matter halo and stellar envelope, and that the remnant nucleus was consistent with an object like NGC~1851. They also showed that mergers of globular clusters in host dwarf galaxies were possible, which in the case of NGC~1851 could explain its multiple stellar populations.

\subsection{Chemistry}
\label{chemistry}
NGC~1851 exhibits the typical GC light element star-to-star abundance dispersions and relations that are believed to result from multiple star formation episodes. Even for GC systems with multiple stellar populations it is unusual in being one of a handful of clusters that show evidence (detailed below) of variations in overall [Fe/H] metallicity and C+N+O abundances. It shows some similarities to $\omega$ Centauri, which exhibits large star-to-star abundance variations in all elements (\citealt{Lee:1999fk} and \citealt{Bedin:2004qy}), but also some similarities to clusters like NGC 6752 that have no [Fe/H] variation but show light element abundance variations \citep{Yong:2013uq}. However, some results have been conflicting, especially the question of whether there is a scatter in the sum of the C+N+O abundances and whether this sum correlates with the metallicities or heavy-element abundances.

Eight bright giant member stars analysed by \citet{Yong:2008lr} (hereafter YG08) showed abundance variations of the s-process elements Zr and La that were correlated with Al and anticorrelated with O. There was a hint that these s-process element abundances clustered around two values, but the sample size was small. An analyis of high-resolution spectroscopic data for four of the YG08 RGB stars was reported in \citet{Yong:2009fk}. These stars showed a large C+N+O abundance variation that correlated with the light elements Na and Al, and also with the s-process elements Zr and La.

The largest extant high resolution spectroscopic study of NGC~1851 stars, \cite{Carretta:2010lr, Carretta:2011lr} (hereafter C10), obtained spectra of 124 RGB stars and found a measurable [Fe/H] spread of 0.06 to 0.08~dex. A division into metal-rich and metal-poor groups based on [Fe/H] and [Ba/H] also shows many differences in various abundance correlations and anti-correlations. They proposed that NGC~1851 is the product of a merger of two globular clusters, with an age difference of $\sim$1 Gyr and different Fe and $\alpha$-capture element abundances, that formed in a now disrupted dwarf galaxy.

There have been several other spectroscopic studies of NGC~1851 stars. \citet{Lardo:2012fj} obtained C and N abundances in 64 SGB and MS stars and found significant anti-correlations in [C/H] and [C/N] and significant variations in C, N and total C+N content when divided into fainter and brighter samples. \citet{Villanova:2010lr} studied a sample of 15 red giant branch (RGB) stars in NGC~1851. Their analysis did not show variations in iron-peak elements or total C+N+O content. In contrast, \citet{Yong:2015fk} found a large ($\sim$0.6~dex) spread in the sum of the C+N+O abundances in AGB and RGB stars, providing support for the idea of coeval SGB populations with different C+N+O content. 

While there has been a large-scale study of C and N in SGB and MS stars \citep{Lardo:2012fj}, and a smaller study in RGB stars \citep{Villanova:2010lr}, C and N abundances have not been measured for a large sample of RGB stars, and not for stars with measured light element abundances. This study was specifically aimed at expanding the set of abundance information available for the RGB stars in the C10 studies. In particular we aimed to carry out a thorough survey of C and N abundances for the same stars, which can be determined from moderate-resolution blue spectra. Combined with the proton-capture (O, Na, Mg, Al), $\alpha$-capture (Si, Ca, Ti) and s-process (Y, Zr, Ba, La) element abundances obtained in their study, this will enable an assessment of whether the C-N and O-Na anticorrelations are tied tightly together in NGC~1851, or whether the relative contributions of the different proton-capture cycles to cluster self-enrichment can vary. In addition, we expect to settle the questions of whether there is a scatter in the sum of the C+N+O abundances and whether the sum correlates with the metallicities or heavy-element abundances determined by C10 (see the conflicting results mentioned above from \citealt{Carretta:2011lr}, \citealt{Yong:2009fk}, \citealt{Villanova:2010lr} and \citealt{Yong:2015fk}). A second paper (in preparation) will address the determination of C and N abundances and discuss these questions.

Because of the large field-of-view of AAOmega, we were also able to investigate the reports detailed above in Subsect. \ref{spatial} that there is an extended halo of stars around NGC~1851. This paper describes the observations and first data reduction and analysis of the sample obtained as well as measurements of $V_r$ for all observed stars, and we assess the evidence for a stellar halo around NGC~1851.

\section{Sample selection, observations and initial data reduction}
\label{obs}

We used the AAOmega spectrograph on the Anglo-Australian Telescope (AAT) at Siding Spring Observatory in two degree field (2dF) fibre positioner multi-object spectrograph (MOS) mode to obtain the spectra\footnote{Manuals and technical details at \url{http/www.aao.gov.au/2dF/aaomega}}. The target list was made up of the 124 RGB stars from the C10 study and four of the giant stars analysed by YG08 (the main targets of our study), plus a third field sample made up of stars with similar 2MASS \citet{2006AJ....131.1163S} photometry to that of NGC~1851 RGB stars. The latter were selected with the aim of allowing for serendipitous discovery of new cluster members, extratidal cluster halo stars or of field stars with unusual abundance patterns. Observing details are listed in Table \ref{tab:tab001}. 

\begin{table*}
\begin{minipage}{100mm}
 \centering
    \caption{Observing details.}
    \label{tab:tab001}
    \begin{tabular}{c c c c c c c}
    \hline
    Date             & Start Time    & Field name & Magnitude          & Exposures           & Seeing      & Airmass     \\
                         & (UTC)    &                          & limits                   & (number x s)       & (arcsec)     & (mid exposure)         \\[0.5ex]
    \hline
    2012 Dec 17 & 11:30   &  bright-f1           & $V<14.5$             &  3 x 540               & 1.4            & 1.10           \\
    2012 Dec 17 & 12:30   &  faint-f1-long     & $V>14.5$             &  2 x 2700             & 1.4            & 1.02           \\
    2012 Dec 17 & 14:21   &  faint-f1-short    & $V>14.5$             &  1 x 2700             & 1.3            & 1.05           \\
    2012 Dec 17 & 15:44   &  faint-f2-long     & $V>14.5$             &  2 x 2700             & 1.5            & 1.11            \\
    2012 Dec 18 & 10:51   &  faint-f2-short    & $V>14.5$             &  1 x 2700             & 1.6            & 1.14            \\
    2012 Dec 18 & 15:34   &  faint-f3-long     & $V>14.5$             &   2 x 2700            & 1.4            & 1.24            \\[1ex]
    \hline 
    \end{tabular}
\end{minipage}
\end{table*}

The red channel spectra were used for determination of stellar heliocentric radial velocities ($V_r$) and metallicities [Fe/H]. We used the standard 2dF data reduction (2dFdr) pipeline, as well as the \textsc{IRAF} task \texttt{splot} to remove skylines.

\begin{figure}
\includegraphics[width=84mm]{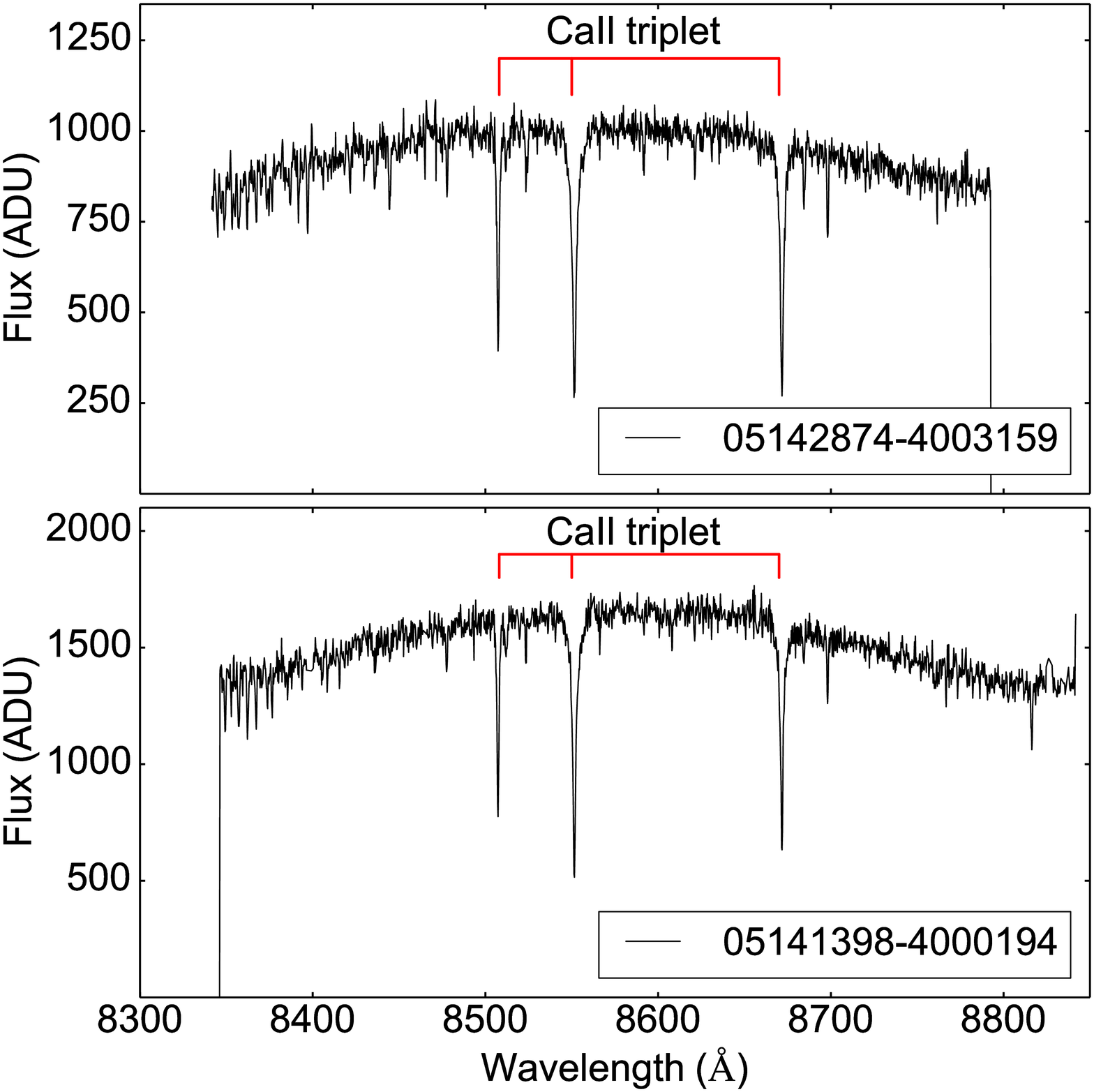}
\caption{Example spectra. Top panel: red channel spectrum of the star used as the template for \texttt{fxcor} $V_r$ measurements - C10 Simbad reference [CLG2011] 30965; 2MASS identification 05142874-4003159. Bottom panel: red channel spectrum of one of our probable new member stars - 2MASS identification 05141398-4000194. The Calcium II triplet absorption lines are indicated. Note: flux scales are different on each plot.}
\label{example_spectra}
\end{figure}

A total of 1876 spectra were obtained. Spectra with low signal-to-noise ratio (SNR) and/or bad columns or other artefacts were discarded leaving a total of 1730 spectra for the analysis. Because of the selection algorithm of the 2dF configuration software, 541 stars were observed more than once in different pointings. In total the useable spectra of 1149 unique stars were obtained, including 106 stars from the C10 study and the four stars from YG08. In the vicinity of the CaT in the red channel, the median SNR of spectra obtained was $\sim$28. Spectra of the star used as a template for calculation of $V_r$ and of one of our probable new member stars are shown in Fig. \ref{example_spectra} to illustrate the quality of our data and the similarity of the spectra in our observations.

\section{Heliocentric radial velocities}
\label{vr}

We used the \textsc{IRAF} task \texttt{fxcor} to derive the $V_r$ of observed stars using the C10 star [CLG2011] 30965 (2MASS ID 05142874-4003159) in our sample as a template star. Mean $V_r$ values were calculated for the 541 stars with multiple observations by weighting their \texttt{fxcor} $V_r$ values with their respective \texttt{fxcor} $V_r$ errors. 

\begin{figure}
\includegraphics[width=84mm]{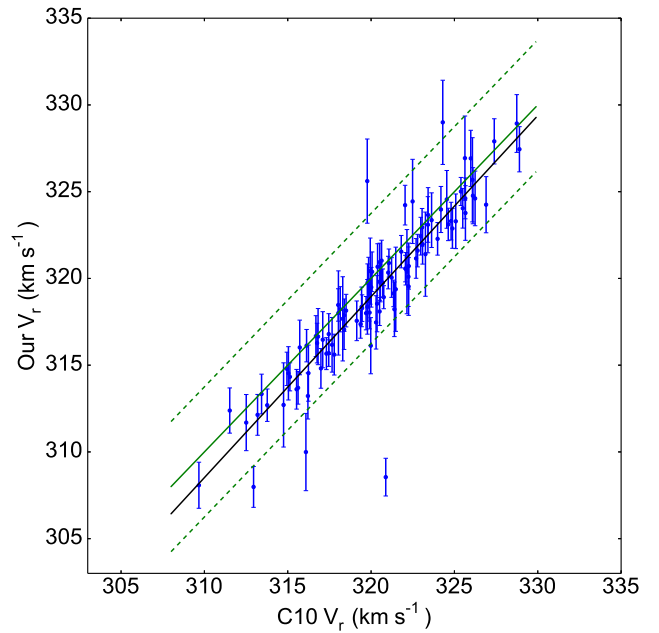}
\caption{$V_r$ of all observed C10 stars compared to our $V_r$ measurements of the same stars. The solid black line is the linear best fit to our data, the solid green line shows the one-to-one relationship and the dashed green lines show $\pm$1$\sigma$ of the one-to-one relationship.}
\label{vr_vr}
\end{figure}
 
Multiple observations of the same target were also used to estimate the $V_r$ errors for stars. An estimate of the standard deviation of $V_r$ ($\sigma_v$) was calculated using the simplified statistics for small number of observations methods of \citet{Dixon:1951bk}. These $\sigma_v$ for multiple observations were then compiled as a function of the median continuum level of the spectrum between the two strongest CaT lines. The values of the mean $\sigma_v$ for each bin for stars with multiple observations were then used as the errors $\Delta V_r$ for stars with single observations as a function of their continuum level. For stars with $N$ observations, $\Delta V_r$ was calculated by dividing $\sigma_v$ by $\sqrt{N}$.

\begin{figure*}
\flushleft{\includegraphics[scale=0.55]{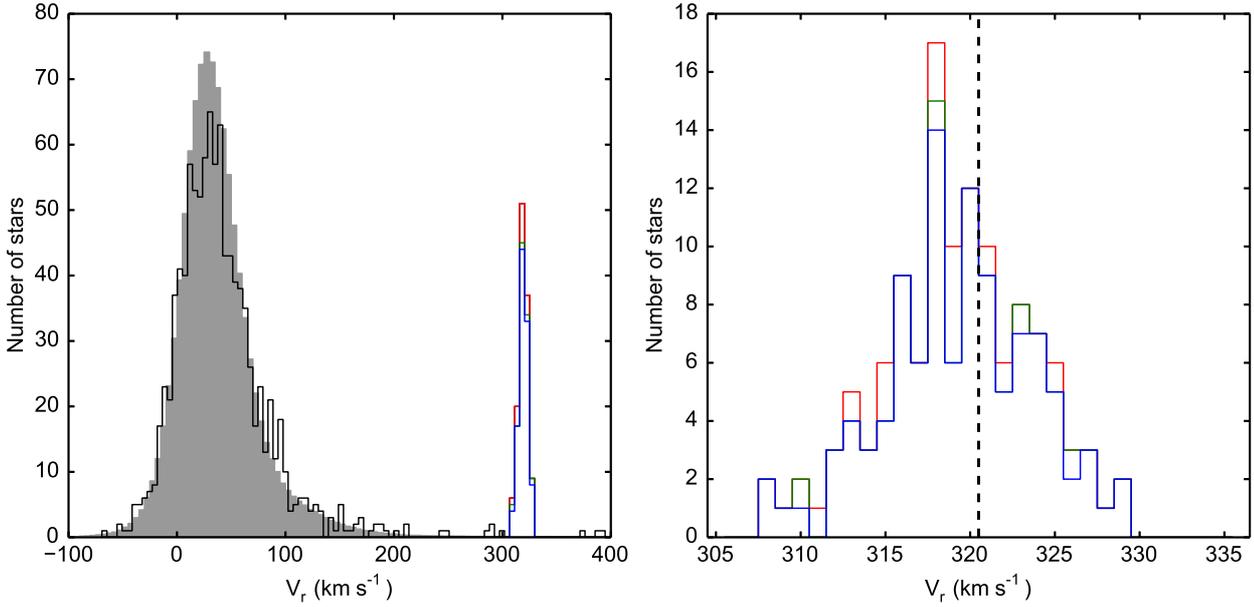}}
\caption{Left panel: $V_r$ distribution of all observed stars with the predicted $V_r$ distribution of the Milky Way model generated by the Galaxia code \citep{Sharma:2011lr} overplotted as a grey histogram. Right panel: Expanded $V_r$ distribution of cluster member and cluster halo stars. Blue denotes known C10 members, green known YG08 members and red are stars that are probable new cluster members or cluster halo stars. The vertical dashed line shows the \citealt{Harris:1996lr} (2010 edition) catalogue $V_r$ of 320.5~\kms.}
\label{vr_hist}
\end{figure*}

We compared our $V_r$ values for the stars in common with the C10 study. Fig. \ref{vr_vr} shows the C10 $V_r$ of all 106 stars recovered in our study compared to our $V_r$ measurements of the same stars. Six of the 106 stars (5.6~per~cent) had measured $V_r$ more than $\pm$1$\sigma$ different from those listed in C10. Binarity is a possible explanation for these differences - binary stars will have variable $V_r$, but will likely still have single-line spectra since our targets are red giant stars. The fraction of stars with $V_r$ more than one $\pm$1$\sigma$ different is consistent with the estimated binary fraction of stars in globular clusters (2~per~cent for the most massive clusters to 20~per~cent for the less massive ones, with a mass weighted average of $\sim$5~per~cent, \citealt{Milone:2008fj}). It is also worth noting that three of these stars are among the fainter stars in our sample, so measurement of their $V_r$ may be affected by the lower SNR of their spectra rather than by binarity.

\begin{table*}
\begin{minipage}{170mm}
 \centering
    \caption{Probable members based on $V_r$. [Fe/H] values marked with * are derived using a $V$ magnitude estimated from 2MASS $J$  and  $K_s$ magnitudes. R = radial distance from cluster centre. Status: m = probable new cluster member star, h = probable extratidal cluster halo star.}
    \label{tab002}
    \begin{tabular}{c c c c c c c c c c c}
    \hline
    2MASS                & RA               & Dec                & $V_r$   & $\Delta V_r$   & [Fe/H]    & $\Delta$[Fe/H] & $\mu_{\alpha}cos(\delta)$   & $\mu_{\delta}$    & R                   & Status \\
    ID                        &                     &                        &  (\kms)  &  (\kms)            & (dex)     & (dex)                & (mas~yr$^{-1}$)                   & (mas~yr$^{-1}$)  & (arcmin)        &             \\ [0.5ex]
    \hline
05111553-4018102  &  5:11:15.54  &  -40:18:10.25  &  318.7  &  2.7                 &  -1.27    &  0.09                & -0.1                                      & 0.7                       & 36.1              &   h    \\
05135811-3956592  &  5:13:58.11  &  -39:56:59.25  &  324.7  &  1.1                 &  -1.38*   &  0.06                & -1.3  				      & -0.1  		    & 6.0  		   &   m   \\
05140637-4024514  &  5:14:06.37  &  -40:24:51.40  &  321.3  &  1.6                &  -1.02*   &  0.05                & -8.5  				      & -2.9  		    & 22.1  		   &   h    \\
05140736-4000020  &  5:14:07.36  &  -40:00:02.04  &  312.7  &  1.6                &  -1.15*   &  0.05                & -10.8    				      & 13.0  		    & 2.8   		   &   m   \\
05141174-4002180  &  5:14:11.74  &  -40:02:18.07  &  319.0  &  1.6                 &  -0.90*   &  0.03               & -  					      & -  		    	    & 1.1  		   &   m   \\
05141359-4001109  &  5:14:13.59  &  -40:01:10.98  &  314.9  &  1.6                 &  -1.56*   &  0.07               & -0.3  				      & 4.7  			    & 2.1  		   &   m   \\
05141398-4000194  &  5:14:13.99  &  -40:00:19.39  &  321.6  &  1.6                 &  -1.44    &  0.06               & -1.7  				      & 12.1  		    & 2.8  		   &   m   \\
05142006-4005116  &  5:14:20.06  &  -40:05:11.69  &  315.2  &  2.2                 &  -1.19    &  0.06                & 3.3  				      & 0.3  			    & 3.5  		   &   m   \\
05142069-3958560  &  5:14:20.70  &  -39:58:56.02  &  311.2  &  1.6                 &  -1.29    &  0.05                & 4.7  				      & -0.7  		    & 4.7  		   &   m   \\
05142088-4005477  &  5:14:20.89  &  -40:05:47.74  &  318.8  &  1.3                 &  -1.21*  &  0.05                & -2.1  				      & 2.1  			    & 4.0  		   &   m   \\
05142994-4000251  &  5:14:29.95  &  -40:00:25.08  &  318.3  &  1.3                 &  -1.08    &  0.04                & -4.6 				      & 1.3  			    & 5.0  		   &   m   \\
05153906-3947485  &  5:15:39.06  &  -39:47:48.52  &  318.6  &  1.3                 &  -1.08    &  0.04                & 0.2  				      & -1.9  		    & 23.2  		   &   h    \\
05162438-4019326  &  5:16:24.38  &  -40:19:32.62  &  317.9  &  1.1                 &  -1.10    &  0.04                & -1.5  				      & -0.4  		    & 31.2  		   &   h   \\  \\[1ex]    
    \hline 
    \end{tabular}
\end{minipage}
\end{table*}

Selection as a probable new member was based on the $V_r$ range of the recovered C10 and YG08 groups. The minimum and maximum measured $V_r$ for stars in these groups was 308.0~\kms~and 328.9~\kms~respectively. We found 13 stars in our field sample with $V_r$ within this range. 

Fig. \ref{vr_hist} shows the $V_r$ distribution of observed stars. The left panel shows all the observed stars, with the predicted $V_r$ distribution of the Milky Way model generated by the Galaxia code (\citet{Sharma:2011lr}, see Sect. \ref{gal}) overplotted as a grey histogram. The distribution shows two prominent peaks: i) a peak with a mean $V_r \sim$40~\kms~consisting of 1015 stars with a spread in radial velocity from -100~\kms $<V_r<$ 200~\kms~that we identify as Galactic disc or halo field stars (non cluster members), and ii) a peak with a mean $V_r \sim$320\kms~containing 123 stars that we identify as members of NGC~1851. In the following we will refer to stars in the first group as non-members, stars in the second group inside the tidal radius as cluster stars or members while stars outside the tidal radius will be referred to as cluster halo stars. In the second group, we recover 106 of 124 of the red giant cluster members in the C10 group and the four red giant cluster members observed in the YG08 group. We also identify 13 probable new cluster members or cluster halo stars of NGC~1851. The right panel shows an expanded plot of the $V_r$ distribution, showing all the observed cluster member and cluster halo stars and a vertical line showing the $V_r$ of NGC~1851. As well as the two peaks we also observe a small number of stars with $V_r$ values intermediate between the two peaks (eight between 200~\kms~and the lower $V_r$ membership limit of 308~\kms) and three stars with very high radial velocities of 371.8, 388.8 and 394.8~\kms. These stars are discussed further in Sect. \ref{high_Vr_stars}.

The mean and standard deviation of the $V_r$ distribution of cluster members and cluster halo stars is 319.3~\kms~and 4.4~\kms, which is similar to previous estimates for NGC~1851 (320.5 \error\ 0.6~\kms~from \citealt{Harris:1996lr} (2010 edition). The $V_r$ of these stars strongly suggests that all these targets are likely cluster members or cluster halo stars, and, where available, this is supported by their positions, proper motions, metallicities (Sect. \ref{metal}) and positions on the colour-magnitude diagrams (Sect. \ref{cmd}). Details of these stars are listed in Table \ref{tab002}.

There are no indications of an overdensity at $\sim$180~\kms~in our observed distribution that might correspond to the tidal tail at this $V_r$ reported by \citet{Sollima:2012uq}. However the number of stars for comparison is small, especially if we exclude the number of stars that are predicted by the Milky Way model generated by the Galaxia code distribution to be in this range of $V_r$. Therefore our observations do not provide additional support for the existence of a stellar stream at $\sim$180~\kms.

\section{Spatial distribution and proper motions}
\label{sdpm}

\begin{figure*}
\flushleft{\includegraphics[scale=0.55]{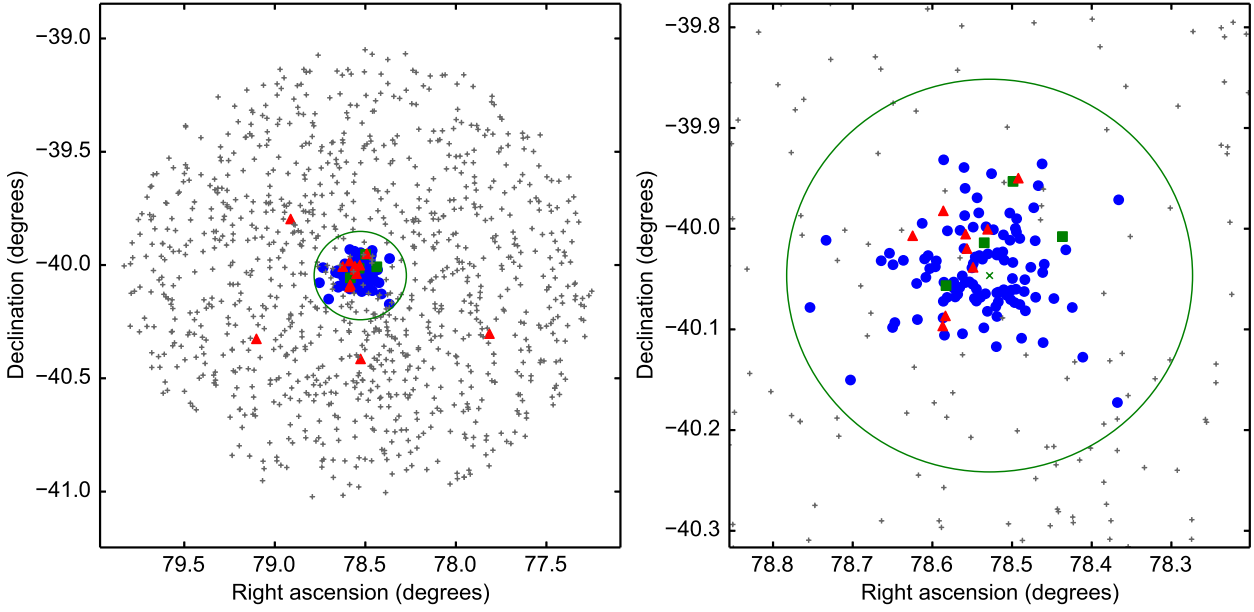}}
\caption{Left panel: Spatial distribution of observed stars for the complete 2\degr~field of 2dF/AAOmega. Right panel: Expanded version showing the observed stars inside the tidal radius. The green cross and circle show the \citealt{Harris:1996lr} (2010 edition) catalogue centre position and the tidal radius of 11.7~arcmin. Grey crosses show non members, blue circles denote known C10 members, green squares are known YG08 members and red triangles are probable new cluster members or cluster halo stars.}
\label{RADec}
\end{figure*}

Figure \ref{RADec} shows the spatial (RA and Dec) distribution of observed stars. Of the 13 stars classified as probable new cluster members or cluster halo stars on the basis of $V_r$, nine are within our adopted \citet{Carballo-Bello:2012lm} tidal radius of 11.7~arcmin and are identified as probable current members of NGC~1851. A further four are outside at distances up to $\sim$3.1 times the adopted tidal radius. Members of a stellar halo of NGC~1851 are expected to share its distinctive $V_r$ signature, so these are identified as probable extratidal cluster halo stars. The identification of these cluster halo stars provides further evidence of the existence of a stellar halo surrounding NGC~1851. Details of these stars are listed in Table \ref{tab002}.

The absolute proper motions of 106 of the C10 stars, nine out of ten of the probable new cluster members and all four probable new cluster halo stars of NGC~1851 were available in the UCAC4 catalogue \citep{Zacharias:2013lr}. The proper motion of NGC~1851 ($\mu_{\alpha}cos(\delta)\ = 1.28\pm0.68~mas~yr^{-1}$, $\mu_{\delta}\ = 2.39\pm0.65~mas~yr^{-1}$) was obtained from \citet{Dinescu:1999lr}. Figure \ref{pm} shows a plot of the absolute proper motions of the observed C10 stars (only 20 stars are plotted to avoid overcrowding) and 12 of the the 13 probable new cluster members or cluster halo stars of NGC~1851. The proper motion of NGC~1851 is shown by the intersection of the dashed lines. It is interesting to note that 15 of the 106 C10 stars have UCAC4 catalogue proper motions in RA and/or Dec greater than $50~mas~yr^{-1}$ and 34 have proper motions in RA and/or Dec greater than $25~mas~yr^{-1}$. This scatter in proper motion values may simply reflect the difficulty of making proper motion measurements in crowded stellar fields. The proper motions of the probable new cluster members and cluster halo stars are listed in Table \ref{tab002} and all are consistent with the proper motion of NGC~1851.

\begin{figure}
\includegraphics[width=84mm]{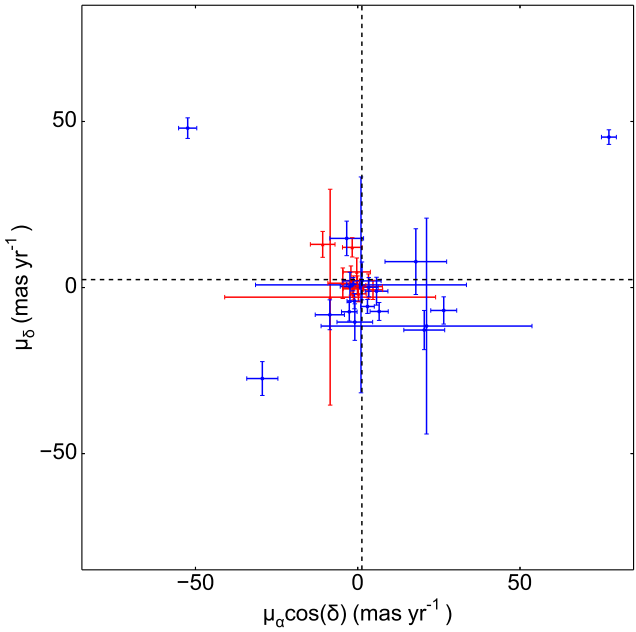}
\caption{Absolute proper motions of 20 of the observed C10 stars (blue) and 12 of the probable new cluster members or cluster halo stars of NGC~1851 (red). The proper motion of NGC~1851 is shown by the intersection of the dashed lines.}
\label{pm}
\end{figure}

\section{Metallicity from the C\lowercase{a} triplet lines}
\label{metal}

Applying the method of \citet{Armandroff:1991fk} we used measurements of the equivalent widths ($EW$) of the CaT lines to calculate stellar metallicity ([Fe/H]). A number of CaT empirical relations have been proposed by different authors. They connect a linear combination of the equivalent widths of the CaT lines and the luminosity of the star (usually the $V$ magnitude of the star relative to the HB of the cluster: $V-V_{HB}$) to its [Fe/H] value. 

We used the \textsc{IRAF} task splot to measure the $EW$s of the CaT lines (8498~\AA $\Rightarrow EW_{8498}$, 8542~\AA $\Rightarrow EW_{8542}$ and 8662~\AA $\Rightarrow EW_{8662}$) in the red channel spectra of the known C10 and YG08 members plus the probable new cluster members and cluster halo stars. Gaussian profiles were fitted to these lines to measure the EWs. 

The calibration procedure is valid for $V-V_{HB}$ brighter than $\sim$0.0 and for a range $\sim$2 in $V-V_{HB}$ luminosities. To include all 13 of the probable new cluster members and cluster halo stars, a faint limit of $V-V_{HB}<0.3$ and a bright limit of $-1.7<V-V_{HB}$ were chosen. These criteria were also applied to the C10 and YG08 stars to select a comparison sample. 64 of the 106 stars in the C10 group matched these criteria, but the YG08 stars are somewhat brighter and did not fit within the limits. 

Fig. \ref{V_VHB_summed_EW_scatter} shows the sum of the measured $EW$s of the CaT lines ($EW_{8542} + EW_{8662}$) of the 13 new cluster members and cluster halo stars plus the 64 C10 stars, plotted against the magnitude difference from the horizontal branch $V-V_{HB}$.

\begin{figure}
\includegraphics[width=84mm]{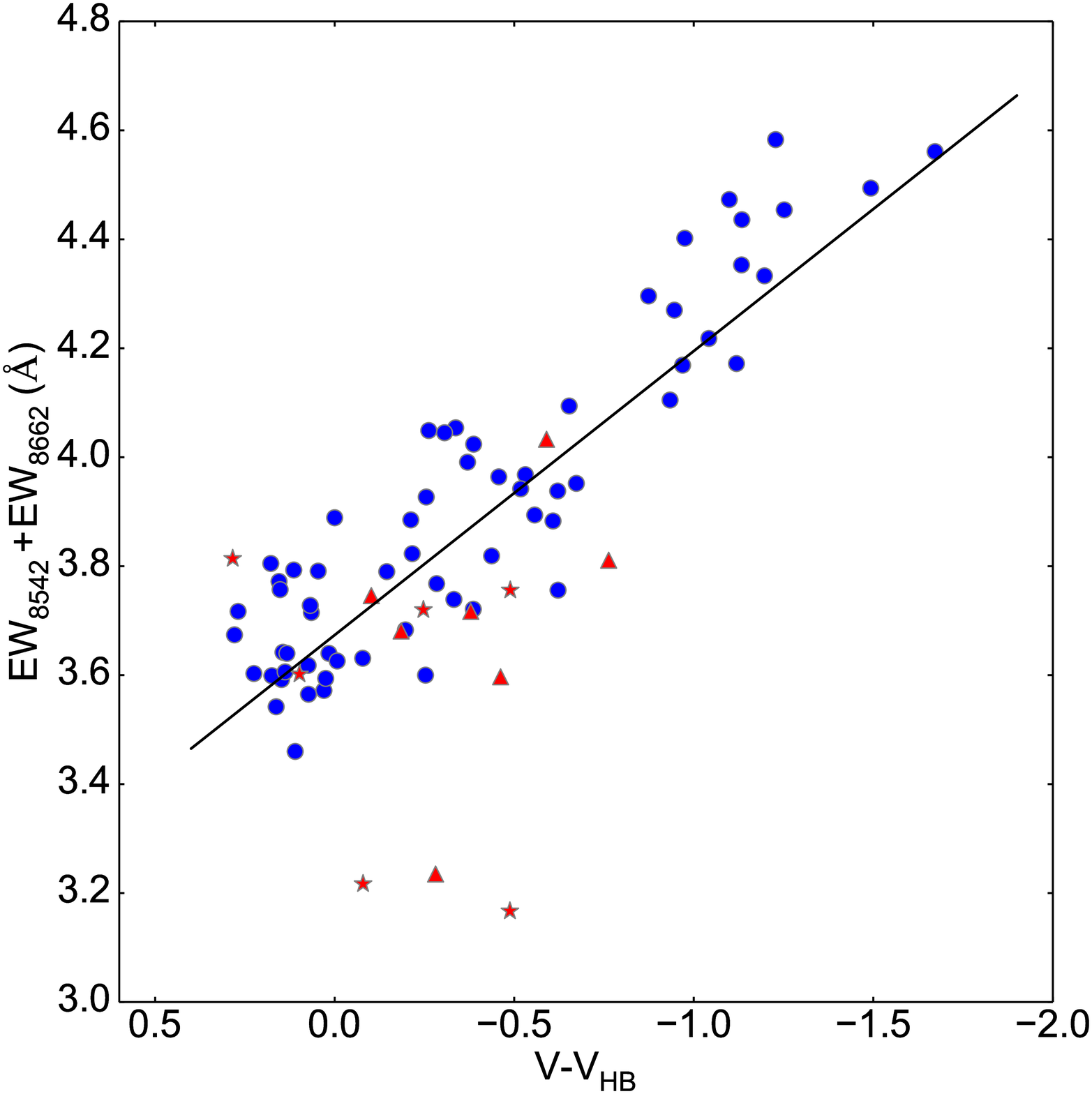}
\caption{The sum of the $EW$s of the CaT lines plotted against the magnitude difference from the horizontal branch $V-V_{HB}$. Blue circles indicate known C10 cluster members, red triangles are probable new cluster members and cluster halo stars with V magnitudes from APASS and red stars are probable new cluster members and cluster halo stars with V magnitudes estimated from J and K magnitudes.}
\label{V_VHB_summed_EW_scatter}
\end{figure}

To calibrate the EWs in terms of [Fe/H] we adopted the empirical calibration equation of \citet{Yong:2014lr}: 
\begin{equation}
\mathrm{[Fe/H]} = 0.59(\pm0.04)W' - 2.81(\pm0.16)
\end{equation}
where the reduced equivalent width $W'$ is :
\begin{equation}
W' = EW_{8542} + EW_{8662} + 0.60(\pm0.02)(V-V_{HB})
\end{equation}

$V$ magnitudes were listed in the catalogue for all 64 of the C10 stars. $V$ magnitudes were available from APASS \citep{Henden:2009qy} for seven of the 13 probable new cluster members and cluster halo stars. For the remaining six stars, an estimate $V_{J,K_s}$ was made from 2MASS $J$ and $K_s$ magnitudes. The calibration formula used was that used to calculate $V$ magnitudes for the GALAH survey input catalogue \citep{De-Silva:2015lr}:
\begin{equation}
V_{J,K_s} = K_s + 2(J - K_s + 0.14) + 0.382e^{((J-K_s-0.2)/0.50)}
\end{equation}

There are differences of up to $\pm \sim$0.5 mag when $V$ magnitudes obtained with this calibration are compared with literature values for C10 stars, which translates into an error of $\pm \sim$0.4~dex in [Fe/H]. Given this, the values of [Fe/H] for these stars are best regarded as indicative values only.

Details of the [Fe/H] of the 13 probable new cluster members and cluster halo stars are listed in Table \ref{tab002}, with the [Fe/H] of stars obtained using an estimated $V$ magnitude marked with an asterisk.

Fig. \ref{FeH_hist} shows the [Fe/H] distribution of the 13 probable new cluster members and cluster halo stars and the 64 C10 members of NGC~1851, with the predicted [Fe/H] distribution of the Milky Way model generated by the Galaxia code (\citet{Sharma:2011lr}, see Sect. \ref{gal}) overplotted as a grey histogram. The model distribution is normalised to the number of observed known members (64) plus the number of probable new cluster members and cluster halo stars (13). The distribution shows a prominent peak with a mean [Fe/H] = -1.10 that we identify as members of NGC~1851. This group includes the 64 homogeneously selected RGB stars from the C10 group and the 13 probable new cluster members or cluster halo stars of NGC~1851.

The mean and standard deviation of the observed [Fe/H] distribution are -1.10 and 0.11 respectively. Seven of our probable new cluster members or cluster halo stars lie within $\pm$1$\sigma$ of the [Fe/H] metallicity of NGC~1851 and 10 within $\pm$2$\sigma$, so most of these stars have metallicities that are consistent with that of NGC~1851.

\begin{figure}
\includegraphics[width=84mm]{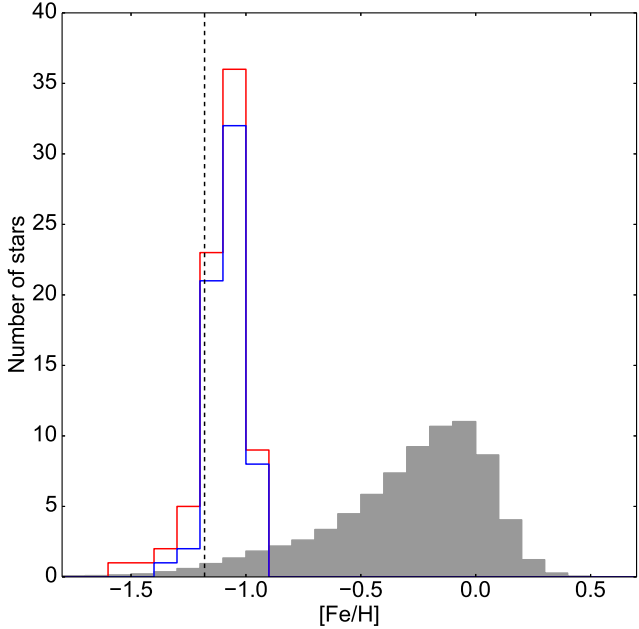}
\caption{[Fe/H] distribution of cluster members and cluster halo stars with the predicted distribution of the Milky Way model generated by the Galaxia code \citep{Sharma:2011lr} overplotted as a grey histogram. Blue indicates known C10 cluster members and red denotes stars that are probable new cluster members and cluster halo stars. The vertical dashed line shows the \citealt{Harris:1996lr} (2010 edition) catalogue [Fe/H] of -1.18.}
\label{FeH_hist}
\end{figure}

\section{Comparison of observed $\mathrm{V_r}$ and [F\lowercase{e}/H] distributions with a Milky way model}
\label{gal}

To investigate further whether the observed $V_r$ and [Fe/H] distributions were associated with NGC~1851, we compared our observations with Milky Way models generated by the Galaxia code \citep{Sharma:2011lr}. 

The Galaxia code was used to generate a synthetic catalogue of stars covering the same area of sky that can be compared with a set of observations. Given colour, magnitude and spatial constraints, Galaxia predicts radial velocities, metallicities and other properties of stars in that area of sky. The Besan\c con Milky Way model \citep{Robin:2003fk} was used for the disk and the simulated N-body models of \citet{Bullock:2005fk} for the stellar halo. To compare our observations with the model the code was run to generate 1000 models covering the 2\degr~field centred on the \citealt{Harris:1996lr} (2010 edition) catalogue centre position of NGC~1851 with no colour or magnitude restrictions. $V_r$ and [Fe/H] distributions were then constructed after applying the same colour and magnitude criteria for selection from the 2MASS catalogue that were used in the target selection process ($0.1<J-H<0.9$, $0<J-K_s<1.2$ and $13<J<14.5$). The $V_r$ distribution was then summed and normalised to have the same number of stars as the number of observed non-member stars, while the [Fe/H] distribution was normalised to have the same number of stars as the number of observed known and probable new cluster members and cluster halo stars. 

In Fig. \ref{vr_hist} and Fig. \ref{FeH_hist} the observed and model $V_r$ and [Fe/H] distributions are plotted. It is immediately apparent that the cluster stars are not reproduced by the models in either the $V_r$ or the [Fe/H] distributions. We then performed a Kolmogorov-Smirnov (K-S) test for the null hypothesis that the model $V_r$ and [Fe/H] distributions and the observed $V_r$ and [Fe/H] distributions are drawn from the same continuous distributions. 

\begin{figure*}
\flushleft{\includegraphics[scale=0.5]{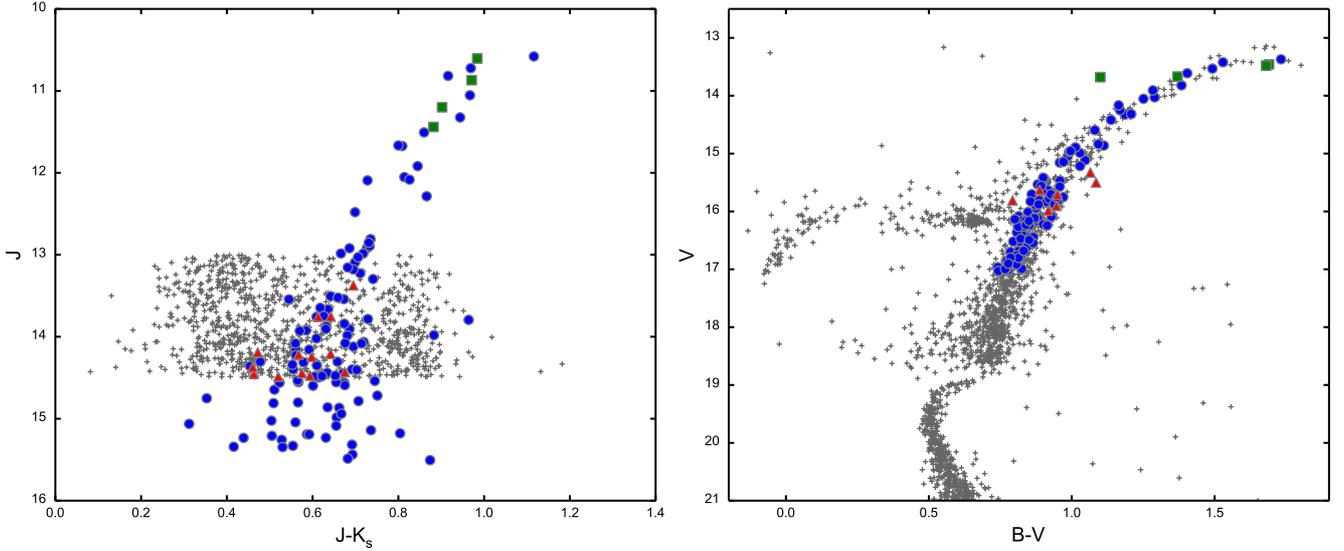}}
\caption{Left panel: 2MASS ($J$ vs $J-K_s$) CMD. The grey crosses show observed non member stars, blue circles indicate known C10 members, green squares known YG08 members and red triangles denote stars that are probable new cluster members and cluster halo stars. Right panel: $V$ vs $B-V$ CMD. Here the grey crosses show \citet{Walker:1992fk} stars, other symbols are the same as for the left panel except that only the seven probable new cluster members and cluster halo stars with directly measured $B$ and $V$ magnitudes are plotted.}
\label{cmd_combined}
\end{figure*}

For the $V_r$ distribution, this was done in the $V_r$ range from 250 to 400~\kms~to compare NGC~1851 member stars with the average Milky Way model distribution. The mean and standard deviation of the observed $V_r$ distribution were 319.3~\kms~and 4.5~\kms~respectively, compared with 306.7~\kms~and 40.1~\kms~for the average model distribution. The $p$ value calculated by comparing the observed and model distributions with the K-S test was 3$\times 10^{-35}$. For [Fe/H] the mean and standard deviation of the observed distribution were -1.108 and 0.097 respectively, compared with -0.309 and 0.373 for the average model distribution. The $p$ value calculated by comparing the distributions with the K-S test was 9.4$\times 10^{-43}$ .

The $p$ values for comparing the observed and model $V_r$ distributions are effectively zero, hence it is unlikely that the observed and predicted $V_r$ samples are drawn from the same distributions. The Milky Way model generated by the Galaxia code predicts just three stars within the same spatial, colour, magnitude and $V_r$ ranges as the observed stars. Similarly, the $p$ value for the [Fe/H] distribution is effectively zero, hence it is unlikely that the observed and predicted [Fe/H] samples are drawn from the same distributions. The model predicts just six stars within the same spatial, colour, magnitude and [Fe/H] ranges as the C10 stars. Most stars in the observed distributions are thus either current members, probable new cluster members or cluster halo stars of NGC~1851.

\section{Colour-magnitude diagrams}
\label{cmd}

Fig. \ref{cmd_combined} shows 2MASS and optical CMDs ($J$ vs $J-K_s$ and $V$ vs $B-V$) of target stars. The left panel clearly shows the 2MASS selection limits of the third group of target stars ($0<J-K_s<1.2$ and $13<J<14.5$). The right panel shows the $V$ vs $B-V$ CMD of all the known C10 and YG08 cluster members and probable new cluster members and cluster halo stars (with directly measured $B$ and $V$ magnitudes). The stars from the photometric study of \citet{Walker:1992fk} are also plotted and the split nature of the HB in the cluster is obvious. The C10 stars, three of the four YG08 stars and seven of the probable new cluster members and cluster halo stars that had directly measured $B$ and $V$ magnitudes all nicely follow the RGB ridge line of the cluster population.

\section{High $\mathrm{V_r}$ stars}
\label{high_Vr_stars}

\begin{table*}
\begin{minipage}{160mm}
 \centering
    \caption{Other stars with high $V_r$. Stars above the line are those with $V_r$ between 200~\kms~and the lower $V_r$ membership limit for NGC~1851 of 308~\kms, those below the line have $V_r$ greater than the $V_r$ of NGC~1851. R = radial distance from cluster centre.}
    \label{tab003}
    \begin{tabular}{c c c c c c c c}
    \hline
    2MASS                 & RA                & Dec                 & $V_r$   & $\Delta V_r$    & $\mu_{\alpha}cos(\delta)$   	& $\mu_{\delta}$     &  R            \\
identification             &                      &                        &  (\kms)  &  (\kms)             & (mas~yr$^{-1}$)                   	& (mas~yr$^{-1}$)   & (arcmin)  \\[0.5ex]
    \hline
05122418-4051595  &  5:12:24.18   &  -40:51:59.53  &  210.3   &  1.6  		  &  9.3					& -18.4  		       &  52.9   	\\
05130199-3931439  &  5:13:02.00   &  -39:31:43.90  &  283.6   &  1.2                  &  13.7					& -1.8		       &  33.5    	\\
05134950-4013007  &  5:13:49.50  &  -40:13:00.77  &  201.5    &  1.1  		  &  35.2					& -23.8  		       & 10.7    	\\
05141567-4001444  &  5:14:15.68  &  -40:01:44.39  &  301.1    &  1.6  	          &  5.9				        & 6.2			       & 2.0    	\\
05152527-4012529  &  5:15:25.27  &  -40:12:52.94  &  243.7    &  1.3  		  &  16.4					& 8.7 		       &18.1    	\\
05162712-3931579  &  5:16:27.12  &  -39:31:57.95  &  213.6    &  0.9                  &  -1.9					& 8.3	  		       & 41.0    	\\
05163977-4039434  &  5:16:39.77  &  -40:39:43.39  &  288.0    &  2.2                  &  60.4					& -0.9		       & 47.0    	\\
05182129-4008551  &  5:18:21.30  &  -40:08:55.15  &  292.4    &  1.3                  &  1.8					& 3.7 		       & 49.1    	\\
\hline
05133314-4100160  &  5:13:33.14  &  -41:00:16.05  &  371.8  &  1.7  		  &  -1.0					& 5.4			        & 57.8   	\\
05135111-3958381  &  5:13:51.11  &  -39:58:38.13  &  388.8  &  1.3  			  &  5.4					& -2.0			& 5.1    	\\
05184308-4027077  &  5:18:43.09  &  -40:27:07.78  &  394.8  &  1.5  		  &  4.2					& 3.4 		         & 58.1    	\\[1ex]    
    \hline 
    \end{tabular}
\end{minipage}
\end{table*}

There are eight stars with $V_r$ between 200~\kms~and the $V_r$ of NGC~1851 as listed in Table \ref{tab003}. These stars are likely not cluster members so their distance and luminosity are unknown. Any $V-V_{HB}$ measurements would be invalid, so metallicities could not be calculated for these stars. Proper motions obtained from the UCAC4 catalogue \citep{Zacharias:2013lr} are also tabulated. The proper motion of NGC~1851 is $\mu_{\alpha}cos(\delta)\ = 1.28\pm0.68~mas~yr^{-1}$, $\mu_{\delta}\ = 2.39\pm0.65~mas~yr^{-1}$) \citep{Dinescu:1999lr}. Several stars have proper motions that are quite different, but as mentioned in Sect \ref{sdpm} there are also significant numbers of C10 stars in the UCAC4 catalogue with proper motions that are considerably different from that of NGC~1851. One star, 2MASS 05141567-4001444, is located just 2~arcmin from the cluster centre, has a proper motion similar to NGC~1851 and a $V_r$ of 301.1~\kms~which is just outside the $V_r$ criteria we adopted for membership, so it could conceivably be a cluster member. If cluster membership is assumed, the [Fe/H] calculated using the methods of Sect \ref{metal} using an estimated $V_{J,K_s}$  from 2MASS $J$ and $K_s$ magnitudes, is -0.41. This is high for a cluster member, even allowing for errors in calculating [Fe/H] with estimated $V$ magnitudes (discussed above in Sect \ref{metal}). Therefore this star was not included in the list of probable members in Table \ref{tab002}.

There are also three stars with $V_r$ $\sim$50 to 75~\kms~higher than that of NGC~1851 (05133314-4100160 at 371.8, 05135111-3958381 at 388.8 and 05184308-4027077 at 394.8~\kms) listed in Table \ref{tab003}. As these stars are also likely not cluster members, metallicities could not be calculated for them. Based on the similarity of their $V_r$ and proper motions we speculate that the stars might be associated. None of the GCs that are also possibly associated with the Canis Major dwarf galaxy (NGC 1904, NGC 2298 and NGC 2808) have a $V_r$ that is consistent with these stars, in fact there are no GCs in that area of the sky that have a similar $V_r$. Without further information, at this stage we tentatively identify all three as high velocity Galactic halo stars.

\section{Summary and conclusions}
\label{summary}

We report the first results of a spectroscopic survey of NGC~1851 and its surroundings, with good quality spectra of 1149 unique stars. Based on $V_r$ obtained from Fourier cross-correlation of the wavelength range around the near infrared calcium triplet absorption lines, we identify
\begin{enumerate}
\renewcommand{\theenumi}{(\arabic{enumi})}
\item nine probable new cluster members inside the tidal radius
\item four probable extratidal cluster halo stars at distances up to $\sim$3.1 times the tidal radius
\item 106 red giant cluster stars in the \cite{Carretta:2010lr, Carretta:2011lr} study
\item four red giant cluster stars observed in the \citet{Yong:2008lr} study
\item eight stars with $V_r$ values intermediate between those of the Galactic field stars and NGC~1851
\item three stars with very high radial velocities of 371.8, 388.8 and 394.8~\kms, and
\item 1015 non-member Galactic disc or halo field stars with -100~\kms $<V_r<$ 200~\kms
\end{enumerate}

Proper motions were available for eight of the nine probable new cluster members and the four probable extratidal cluster halo stars and all are consistent with that of NGC~1851. The four extratidal cluster halo stars provide further confirmation of the findings of \citet{Olszewski:2009lr}, \citet{Carballo-Bello:2010lr}, \citet{Marino:2014lr} and \citet{Kunder:2014lr} that NGC~1851 is surrounded by a stellar halo system and does not have a classical tidally-limited profile. Apart from the cluster and halo stars our observed radial velocity distribution agrees with the expected distribution from the Besan\c con disk/N-body stellar halo Milky Way model generated by the Galaxia code, and the probability that the $V_r$ distribution of the cluster and halo stars are drawn from the same distribution as the model is effectively zero. In particular there are no indications of an overdensity at $\sim$180~\kms~in our observed distribution that might correspond to the tidal tail at this $V_r$ reported by \citet{Sollima:2012uq}, however the number of observed stars in our sample with comparable $V_r$ is small. 

The metallicities of the \cite{Carretta:2010lr, Carretta:2011lr} stars and the probable new cluster members and cluster halo stars were estimated using equivalent width measurements of the calcium triplet absorption lines. Within the limitations of this method discussed above, the metallicities of most of the probable new cluster members and cluster halo stars were found to be consistent with that of NGC~1851. As with the $V_r$ distribution, the probability that the [Fe/H] distribution of the cluster and halo stars are drawn from the same distribution as the model is effectively zero.

We also identified three stars with unusually high $V_r$, $\sim$50 to 75~\kms~higher than that of NGC~1851. As the velocity separation is large it is likely that they are not associated with NGC~1851. Based on the similarity of their $V_r$ and proper motions we speculate that the stars might be associated and tentatively identify them as high velocity Galactic halo stars.

In our second paper (in preparation) we will address the determination of C and N abundances using the blue channel of the spectra and discuss the questions of the relationship of the C-N and O-Na anticorrelations, whether there is a scatter in the sum of the C+N+O abundances and whether the sum correlates with metallicities or heavy-element abundances.

\section*{Acknowledgements}
\label{ack}
We thank Pete Kuzma, Sanjib Sharma and David Yong for their assistance and helpful discussions, and also Daniela Carollo as one of the initiators of the project. The authors also thank the reviewer for their useful comments and suggestions for improvement. We gratefully acknowledge the support of the Australian Astronomical Observatory. \textsc{IRAF} is distributed by the National Optical Astronomy Observatory, which is operated by the Association of Universities for Research in Astronomy, Inc., under cooperative agreement with the National Science Foundation. This research has made use of NASA's Astrophysics Data System Bibliographic Services. This publication makes use of data products from the Two Micron All Sky Survey, which is a joint project of the University of Massachusetts and the Infrared Processing and Analysis Center/California Institute of Technology, funded by the National Aeronautics and Space Administration and the National Science Foundation. This research has made use of the VizieR catalogue access tool, CDS, Strasbourg, France. This research made use of Astropy, a community-developed core Python package for Astronomy (Astropy Collaboration, 2013).




\bibliographystyle{mnras}
\bibliography{Galactic_archaeology}


\bsp	
\label{lastpage}
\end{document}